\documentclass[aps,floatfix,prd,showpacs]{revtex4}
\usepackage{graphicx}% Include figure files
\usepackage{dcolumn}% Align table columns on decimal point
\usepackage{bm}% bold math

\begin{document}

\def\be{\begin{equation}}
\def\ee{\end{equation}}

\title{Endemic Infrared Divergences in QED3 at Finite Temperature}
\author{Pok Man Lo and Eric S. Swanson}
\affiliation{
Department of Physics and Astronomy, 
University of Pittsburgh, 
Pittsburgh, PA 15260, 
USA.}

\date{\today}

\begin{abstract}
We demonstrate that massless QED in three dimensions contains endemic infrared divergences. It is argued that these divergences do not affect observables; furthermore,  it is possible to choose a gauge that renders the theory finite.
\end{abstract}

\maketitle

\section{Introduction}

Massless field theories in reduced dimensions are often burdened  with infrared singularities that make computation and interpretation difficult. The problem is particularly acute in gauge theories, where mass, and hence an infrared cutoff, cannot be generated in perturbation theory. It is not unusual for these divergences to appear at a given order in perturbation theory, but to disappear once appropriate diagrams are summed. For example the famed resolution of the infrared divergence in the degenerate electron gas relies on summing infinitely many ring diagrams\cite{deg}. 

Another example is provided by massless QED in three dimensions (QED3). The perturbative vacuum polarisation form factor satisfies 
$\Pi(p^2) \sim e^2 |p|,$
thus the $O(e^4)$ fermion bubble insertion correction to the fermion self-energy behaves as 

\be
\int d^3 q\,\frac{1}{q^2} e^2 |q| \frac{1}{q^2} \sim \log \Lambda_{IR}
\ee
where $\Lambda_{IR}$ is an infrared cutoff. 
This divergence is cancelled by a nonanalytic $e^4 \log e^2$, rendering the theory well-behaved\cite{JT}.

%The physical mechanism underlying Coleman's theorem provides another example. The theorem states that two-dimensional field theories do not spontaneously break continuous symmetries. The physical interpretation is that infrared enhancements in quantum fluctuations are sufficient to overwhelm an putative long range order.
%
Finally, we recall that finite temperature QCD in four dimensions suffers from the Linde problem, wherein infrared divergences appear at order $g^6$ in gluonic loop diagrams\cite{linde}. The resolution of this problem is not known.

A variety of novel features and applications have spurred interest in QED3. For example a Chern-Simons-like photon mass is possible in three dimensions\cite{early, DJT}
\be
{\cal L}_{\rm CS} = \mu\frac{1}{4}\epsilon_{\mu\nu\alpha}F^{\mu\nu}A^\alpha.
\ee
This term breaks P and T invariance. It also generates fermion mass at one-loop in perturbation theory.

If one considers the case of $N_f$ massless two-component fermions then it has been argued that chiral symmetry is indeed broken but that the vacuum remains invariant under parity\cite{appel}. This has recently been supported by a numerical solution of the truncated Schwinger-Dyson equations\cite{LS}.

Interestingly, it appears that fermion screening at $N_f \agt 3.3$ is sufficient to restore a chirally symmetric vacuum\cite{chiral}.
Thus this theory illustrates how large mass hierarchies can be dynamically generated, which is of interest to BSM physics\cite{mass-scales}.

QED3 is popular as a model field theory for three dimensional condensed matter systems. Examples include applications to high $T_c$ superconductors, where the relevant 
dynamics is thought to be isolated to copper-oxygen planes in the cuprate\cite{highTc}. It is also considered as a gauge formulation of the 2+1 dimensional Heisenberg spin model\cite{richert} and as a possible model for graphene\cite{graph} and quantum versions of spin-ice\cite{spin-ice}. Naturally, the behaviour of the theory at finite temperature and density is of great interest in these applications.

Infrared divergences are exacerbated  at nonzero temperature because perturbative diagrams are dominated in the infrared limit by the lowest available Matsubara frequency, which is zero in bosonic sums. Thus, even though QED3 is infrared finite at zero temperature, problems may arise again at nonzero temperature. This issue has engendered some confusion in the literature. 
Some authors  have noted that an infrared divergence exists, but have ignored it\cite{ignore}, or imposed an infrared cutoff\cite{cutoff}, or assumed that higher order corrections remove the divergence\cite{higher-order}.
Many authors simply evade the issue entirely by employing the approximation\cite{D00}

\be
iD_{\mu\nu}(\omega,\vec q) \to iD_{00}(0,\vec q).
\ee

We shall argue that infrared divergences are endemic to QED3 at finite temperature. Furthermore, the problem is not alleviated by finite fermion masses. Nevertheless, observables are finite and the theory is well-defined.

\section{QED3 Schwinger-Dyson Equations at Finite Temperature}

We first consider massless abelian gauge theory in three dimensions. The Lagrangian is

\be
{\cal L} = -\frac{1}{4}F^2 + \bar \psi(i\rlap{/}\partial + e\rlap{/}A)\psi - \frac{1}{2\xi}(\partial\cdot A)^2.
\ee
The coupling $e^2$ has units of mass and the theory is superrenormalisable. 
In three dimensions it is possible to form two- and four-dimensional representations of the Dirac algebra.  The two-component theory does not have a chiral symmetry and the massive and massless theories have the same symmetry\cite{chiral}. As mentioned in the Introduction, this 
 theory does admit the possibility of spontaneous photon and fermion mass generation -- and hence parity symmetry breaking.  The four-component theory with $N_f$ fermions is $U(2N_f)$ symmetric, and can break to $U(N_f)\times U(N_f)$ if fermion mass is dynamically generated. We will show that infrared divergences occur in both versions.

\begin{figure}[ht]
\includegraphics[width=10cm,angle=0]{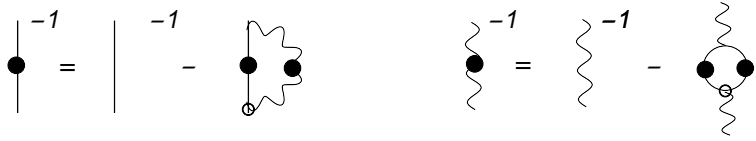}
\caption{Schwinger-Dyson Equations. Solid circles represent full propagators. The open circles represent the full vertex.}
\label{sde-fig}
\end{figure}

The Schwinger-Dyson formalism will be used to verify the occurrence of infrared divergences. The equations for the
fermion and photon propagators are shown in Fig. \ref{sde-fig}. These equations are considered at finite temperature, the extension to finite density is simple and does not change the conclusions. We employ the imaginary time formalism and choose to work covariantly, which necessitates introducing a three-vector, $n^\mu$, that represents the heat bath. Thus the full fermion propagator is

\be
S = \frac{i}{C \rlap{n}/ + A \rlap{/}p - B }.
\label{S-defn-eq}
\ee
Here 
\be
p^\mu = (i\omega_n,\vec p)
\ee
where $\omega_n = (2n+1)\pi T$ is a fermionic Matsubara frequency 
and $A$, $B$, and $C$ are functions of $\omega_n$ and $\vec p^2$.

\subsection{Photon Propagator}

The vacuum polarisation tensor remains transverse at finite temperature, however the presence of an additional three-vector permits two transverse tensors

\be
P^L_{\mu\nu}(n,q) = {\hat q}^\perp_\mu {\hat q}^\perp_\nu
\ee
and
\begin{equation}
P^\perp_{\mu\nu}(n,q) = g_{\mu\nu}-\frac{q_\mu q_\nu}{q^2} - P^L_{\mu\nu}(q).
\end{equation}
A transverse three-vector has been defined as
\be
q^\perp_\mu = q_\mu - n_\mu \frac{q^2}{n\cdot q}.
\ee

%Some useful properties of these tensors are 
%
%\be
%q^\mu P^L_{\mu\nu}(q) = q^\mu P^\perp_{\mu\nu}(q) = 0 \qquad n^\mu P^\perp_{\mu\nu} = 0
%\ee
%
%\begin{equation}
%P^\perp \cdot P^L = 0 \qquad P^\perp \cdot P^\perp = P^\perp \qquad P^L\cdot P^L = P^L.
%\end{equation}
%With $n^\mu = (1,0,0)$ we obtain
%
%\begin{equation}
%P^\perp_{00} = 0 \qquad P^\perp_{0i} = 0 \qquad P^\perp_{ij} = -(\delta_{ij}-\hat q_i \hat q_j).
%\end{equation}
%The minus sign in the last expression is  required in the  covariant normalisation employed here. 

With these definitions one can parameterise the photon self energy as:

\be
\Pi_{\mu\nu}(n, q) = P^\perp_{\mu\nu} \Pi_\perp + P^L_{\mu\nu} \Pi_L
+ i \epsilon_{\mu\nu\alpha} \hat q^\alpha \tilde\Pi + i \hat q^\perp_\mu \epsilon_{\nu\alpha\beta}  {\hat q}^\alpha \hat{q}_\perp^\beta \Pi_4 +  i \hat q^\perp_\nu \epsilon_{\mu\alpha\beta} {\hat q}^\alpha \hat q_\perp^\beta \Pi_5 +
\epsilon_{\mu\alpha\beta}\epsilon_{\nu\alpha'\beta'} {\hat q}^\alpha {\hat q}^{\alpha'} \hat q_\perp^\beta \hat q_\perp^{\beta'}\, \Pi_6
\ee
Note that $\tilde \Pi$, $\Pi_4$, $\Pi_5$, and $\Pi_6$ are all null for four-component fermions and it is possible to combine the $\Pi_4$ and $\Pi_5$ terms into symmetric and antisymmetric tensors.

The full photon propagator is

\be
iD_{\mu\nu}(n, q) = D_\perp P^\perp_{\mu\nu} + D_L P^L_{\mu\nu}  -i\xi \frac{q_\mu q_\nu}{q^4}
+ i \tilde D \epsilon_{\mu\nu\alpha} \hat q^\alpha + i D_4 \hat q^\perp_\mu \epsilon_{\nu\alpha\beta}  {\hat q}^\alpha \hat{q}_\perp^\beta + i D_5 \hat q^\perp_\nu \epsilon_{\mu\alpha\beta} {\hat q}^\alpha \hat q_\perp^\beta + D_6
\epsilon_{\mu\alpha\beta}\epsilon_{\nu\alpha'\beta'} {\hat q}^\alpha {\hat q}^{\alpha'} \hat q_\perp^\beta \hat q_\perp^{\beta'}
\ee

For four-component fermions the form factors are 

\begin{eqnarray}
D_\perp &=& -i \frac{q^2-\Pi_\perp}{(q^2-\Pi_\perp)^2 - \mu^2 q^2} \label{A-4} \\
D_L &=& -i \frac{q^2-\Pi_\perp}{(q^2-\Pi_L)(q^2-\Pi_\perp) - \mu^2q^2} \\
\tilde D &=&  \frac{-i\mu q}{(q^2-\Pi_\perp)^2 - \mu^2 q^2} \\
D_4 &=& \frac{\mu q}{q^2-\Pi_\perp} (D_L-D_\perp) \\
D_5 &=& -D_4 \\
D_6 &=&  \frac{\mu q}{q^2-\Pi_\perp} D_4.
 \label{G-4}
\end{eqnarray}
The form factors are more complicated for two-component fermions;
we consider the case $\mu=0$ and $\tilde\Pi=0$, which is appropriate for isolating infrared divergences. One then obtains

\begin{eqnarray}
D_\perp &=& \frac{-i}{q^2-\Pi_\perp} \label{A-2}\\
D_L &=& -i\frac{q^2-\Pi_\perp - \Pi_6}{(q^2-\Pi_\perp)(q^2-\Pi_L) +\Pi_4\Pi_5 - \Pi_6(q^2-\Pi_L)}\\
\tilde D &=& 0 \\
D_4 &=& -i\frac{\Pi_4}{(q^2-\Pi_\perp)(q^2-\Pi_L) +\Pi_4\Pi_5 - \Pi_6(q^2-\Pi_L)}\\
D_5 &=& -i\frac{\Pi_5}{(q^2-\Pi_\perp)(q^2-\Pi_L) +\Pi_4\Pi_5 - \Pi_6(q^2-\Pi_L)}\\
D_6 &=& -i\frac{\Pi_6(q^2-\Pi_L)-\Pi_4\Pi_5}{(q^2-\Pi_\perp)[(q^2-\Pi_\perp)(q^2-\Pi_L) +\Pi_4\Pi_5 - \Pi_6(q^2-\Pi_L)]}.
 \label{G-2}
\end{eqnarray}

\subsection{Infrared Properties of the Photon Propagator}

We will shortly establish that it is the photon propagator that generates the infrared divergences in QED3 at finite temperature. We therefore examine the properties of the photon form factors in more detail.

At one-loop order, and temporarily generalising to the massive fermion case, one obtains

\be
\Pi_4 =0, \qquad \Pi_5 = 0, \qquad \Pi_6 =0,
\ee
and
\be
\Pi_L^{\rm (mat)}(0,q\to 0) = \frac{2 e^2}{\pi}
\, \left( T \log(1 + {\rm e}^{m/T}) - m/(1+{\rm e}^{-m/T}) \right) 
\equiv m_{\rm el}^2
\ee
which defines the electric screening mass. The superscript indicates that this is the `matter' portion of the form factor, which is defined as the form factor minus its vacuum (zero temperature) value. In the massless limit the screening mass reduces to 
\be
m_{\rm el}^2 = 8\log(2) \alpha T
\ee
with $\alpha = e^2/(4\pi)$. Of course this screening mass regulates infrared divergences for the longitudinal portion of the photon propagator. Unfortunately the same is not true for the transverse ($\Pi_\perp$) and Chern-Simons ($\tilde\Pi$) portions.

At one-loop one obtains
\be
\Pi_\perp^{\rm (mat)}(0,q\to 0) = -8 \alpha \int_m^\infty dE\, \left[ \frac{E^2+m^2}{4 E^2}(\tanh(E/2T)-1) + \frac{E^2-m^2}{8 T E} (1 - \tanh^2(E/2T)) \right] = 0
\ee

The fact that there is no magnetic screening mass at this order is well-known and is true to all orders. The argument\cite{fradkin} employs the exact expression for the photon self-energy:

\be
\Pi_{\mu\nu}(n, q) \sim e^2 T \sum_\nu \int d^2\ell\, {\rm tr}\, [\gamma_\mu S(\ell)\, \Gamma_\nu(q+\ell,\ell) \, S(q+\ell)]
\ee
where $\Gamma_\nu$ is the full three point function.

The Ward identity can be used to simplify this expression for small $q$:
\be
\Gamma_\nu(\ell,\ell) = \frac{\partial S^{-1}(\ell)}{\partial \ell_\nu}.
\ee
Substituting and employing 

\be
\frac{\partial S^{-1}}{\partial \ell_\nu} = - S^{-1} \frac{\partial S}{\partial \ell_\nu} S^{-1}
\ee
then yields

\be
\Pi_{\mu\nu}(0,q\to 0) \sim -e^2 T \sum_\nu \int d^2\ell\, {\rm tr}\, [\gamma_\mu \frac{\partial S}{\partial q_\nu}].
\label{Pi-exact-eq}
\ee
Thus the integral over spatial dimensions vanishes and the magnetic mass must be zero to all orders.

It is possible to enlarge this statement to read

\be
\Pi_\perp(0,q\to 0) = c_\perp q^2 + O(q^4).
\ee
This is explicitly true to $O(e^5)$ in QED4 and Blaizot {\it et al.} argue that it is true to all orders\cite{BI}. 
The basic idea is that the nonvanishing minimum fermionic Matsubara frequency makes the self energy an analytic function of $q^2$. An expansion about $q=0$ then yields $\Pi_\perp \to 0 + O(q^2)$, with the odd terms vanishing due to rotational invariance.
%NOTE: need mel >0 to screen electrostatic internal lines.
We note that this argument generalises directly to three dimensions. This result is important because it implies that there is no dynamical screening in the magnetic sector.

The one-loop expression for the Chern-Simons form factor is

\be
\tilde\Pi(0,q\to 0) = \alpha q\, \tanh(m/2T).
\label{Pi2-temp-eq}
\ee
This result should be compared to the zero temperature form factor

\be
\tilde\Pi(0) =  \alpha q\, \frac{m}{|m|}.
\label{Pi2-Eq}
\ee
One sees that the zero temperature limit correctly recovers Eq. \ref{Pi2-Eq} but as the fermion mass goes to zero $\tilde\Pi(0) \to \pm \alpha q$ whereas the finite temperature version approaches zero if $T$ is kept finite (the correct result is obtained if one takes the limit before performing the integral). Furthermore, when $2N_f$ degenerate fermions are present it is possible to maintain parity invariance of the vacuum (which is expected\cite{VW}) if $N_f$ fermions develop positive dynamical mass and $N_f$ fermions develop negative dynamical mass. Thus it is expected that $\tilde\Pi(0)=0$ and $\tilde\Pi^{\rm (mat)}(0,0) = 0$ on quite general grounds.

A theorem due to Coleman and Hill stipulates that $\tilde\Pi(0)$ receives no further corrections at zero temperature\cite{CH}. The argument considers the effective action obtained by integrating out the fermions;
gauge invariance implies that the leading behaviour of a general $n$-point function is

\be
\Gamma^{(n)}(p_1 \ldots p_n) = O(p_1 p_2).
\ee
Two external photon lines  contribute $O(p^2)$ to any diagram beyond one-loop and so can not contribute to a function that is proportional to $p$. 
Nonzero contributions can only arise from graphs
in which the two external photon lines end on a loop that has no other photon lines attached to it. Thus all
corrections to the photon mass beyond one loop order vanish. Evidently the argument generalises directly to nonzero temperature so that Eq \ref{Pi2-temp-eq} is true to all orders.

Finally, for two-component fermions $\Pi_4$, $\Pi_5$, and $\Pi_6$ need not be zero. However, all three are zero at one-loop order. Furthermore the conditions of the (generalised) Coleman-Hill theorem apply to $\Pi_4$ and $\Pi_5$ implying

\be
\Pi_4(0,q\to 0) = 0, \qquad \Pi_5(0,q\to 0) = 0
\ee
to all orders. Furthermore, as $T \to 0$ one must recover the vacuum form of the photon propagator. Eqs. \ref{A-4}--\ref{G-4} then imply that $\Pi_L(T\to 0; \nu, q) = \Pi_\perp(T\to 0;\nu,  q)$, which is a well-known property of these form factors. Similarly Eqs. \ref{A-2} -- \ref{G-2} imply that

\be
\Pi_4(T\to 0;\nu, q) =0, \qquad \Pi_5(T\to 0; \nu, q) = 0, \qquad \Pi_6(T\to 0; \nu, q) = 0.
\ee

\subsection{Fermion Self-Energy and Infrared Divergences}

The lack of magnetic screening leads directly to an infrared divergence in the fermion self-energy, as we now demonstrate. Consider the exact expression for the fermion self energy that appears in the Schwinger-Dyson equation of Fig. \ref{sde-fig}.

\be
i\Sigma(p) = -ie^2 T \sum_n \int \frac{d^2q}{(2\pi)^2} \gamma_\nu S(q) \Gamma_\mu(p,q) D^{\mu\nu}(p-q).
\ee

We again employ the finite temperature version of the Ward identity to obtain the leading behaviour of the fermion self energy when $q = p -\eta$:

\be
\Gamma_\nu(p+\eta,p) = \frac{\partial S^{-1}(p)}{\partial p_\nu} + \eta^\alpha \frac{\Gamma_\nu}{\partial p_\alpha} + \ldots
\ee
When $n=(1,0,0)$ the leading infrared behaviour is obtained when $\nu$ (or $n\cdot q$) is zero, we therefore set $\nu=0$ in the following. One obtains

\begin{eqnarray}
{\rm div}\, \Sigma(p) &=& -e^2 T [\gamma_\nu S(p) \frac{\partial S^{-1}}{\partial p_\mu}]\, {\rm div}\,  \int_{\Lambda_{IR}} \frac{d^2\eta}{(2\pi)^2}\, D_{\mu\nu}(\eta) \\
&=& e^2 T [\gamma_\nu \frac{\partial S}{\partial p_\mu} S(p)^{-1}]\, {\rm div}\,\int_{\Lambda_{IR}} \frac{d^2\eta}{(2\pi)^2}\,
\left[n_\mu n_\nu D_L + \frac{n_\mu n_\nu - g_{\mu\nu}}{2}(i D_\perp + i D_6  - \frac{\xi}{\eta^2}) \right] 
\label{IR-eq}
\end{eqnarray}
For four-component fermions only the gauge term is infrared divergent if the photon mass is nonzero. If it is zero one has
\be
{\rm div}\, \Sigma(p) = 
  e^2 T [\gamma_\nu \frac{\partial S}{\partial p_\mu} S(p)^{-1}]\, \frac{n_\mu n_\nu - g_{\mu\nu}}{4\pi}\,
\left[\frac{1}{1-c_\perp}  +\xi \right] \log \Lambda_{IR}.
\ee
Similarly for two-component fermions with $\mu=0$ 
\be
{\rm div}\, \Sigma(p) = 
  e^2 T [\gamma_\nu \frac{\partial S}{\partial p_\mu} S(p)^{-1}]\, \frac{n_\mu n_\nu - g_{\mu\nu}}{4\pi}\,
\left[\frac{1}{1-c_\perp} - \frac{c_6}{(1-c_\perp)\,(c_6+c_\perp-1)} + \xi \right] \log \Lambda_{IR}.
\label{c6-eq}
\ee
We have assumed that $\Pi_6(0,q\to 0) \to c_6 q^2$. If there is a screening mass in this form factor one can take the limit as $c_6$ goes to infinity in Eq. \ref{c6-eq}. Furthermore, if parity symmetry is spontaneously broken so that $\tilde \Pi(0) \neq 0$ then only the gauge term has an infrared divergence.

Thus a logarithmic infrared divergence appears in the fermion propagator. This statement is exact, only relying on the Ward identity,the existence of $1/q^2$ terms in the exact photon propagator, and general properties of the photon form factors.
It is clear that a finite fermion mass does not change this conclusion. However, a finite photon mass regulates the transverse part of the propagator, leaving only the divergence in the gauge term.

\section{Discussion and Conclusions}

At first sight the appearance of endemic infrared divergences is problematic; however, we now demonstrate that these divergences do not appear in observables. 
Consider first the case where the 
photon has a mass; the infrared divergence is then associated with the gauge parameter. Thus gauge invariance of observables implies that observables are also independent of the infrared cutoff. 

Furthermore, $S(p)^{-1}$ is zero at the physical pole location, thus 
Eq. \ref{IR-eq} implies that the pole location  is independent of the gauge parameter and the infrared cutoff. 
In fact because the order of the infrared divergence due to the gauge and magnetic portions of the photon propagator is the same and because the prefactors are constants, one can 
simply absorb the magnetic divergence into the gauge parameter, thereby showing that all observables are infrared safe.

Finally, it is always possible to employ the gauge 

\be
\xi = -\frac{1}{1-c_\perp},
\ee
with an analogous expression for two-components fermions, to render all quantities infrared safe.

An all-orders argument for the existence of infrared divergences in the fermion propagator of QED3 at finite temperature has been presented. The argument relies on gauge invariance and analyticity of $n$-point functions at nonzero temperatures. These properties imply that $\tilde\Pi$, $\Pi_\perp$, $\Pi_4$, and $\Pi_5$ all scale as  $O(q^2)$ for small momenta.  The Ward identity can be used to isolate infrared divergences to the full photon propagator, revealing that the lack of magnetic screening in three dimensions leads to a logarithmic infrared divergence unless parity is explicitly broken with a photon mass.

Nevertheless, because the divergences are associated with the exact photon propagator and are of the same form as that arising due to the gauge term, they must cancel in observables. It is therefore possible to define a gauge which eliminates the divergences entirely.

\acknowledgments
This research was supported by the U.S. Department of Energy under contract
DE-FG02-00ER41135 and an Andrew W. Mellon Predoctoral Fellowship.

\end{document}